\documentclass[fleqn,10pt]{wlscirep}
\usepackage[utf8]{inputenc}
\usepackage[T1]{fontenc}
\usepackage{physics}
\usepackage{dsfont}  

\title{Universal Atom Interferometer Simulator - Elastic Scattering Processes}

\author[1,2]{Florian Fitzek}
\author[2,1]{Jan-Niclas Siem\ss}
\author[1]{Stefan Seckmeyer}
\author[1]{Holger Ahlers}
\author[1]{Ernst M.\@ Rasel}
\author[2]{Klemens Hammerer}
\author[1]{Naceur Gaaloul}
\affil[1]{Institut für Quantenoptik, Leibniz Universität Hannover, Welfengarten 1, D-30167, Hannover, Germany}
\affil[2]{Institut für Theoretische Physik, Leibniz Universität Hannover, Appelstra\ss e 2, D-30167, Hannover, Germany}

\affil[*]{gaaloul@iqo.uni-hannover.de}

\begin{abstract}

In this article, we introduce a universal simulator covering all regimes of matter wave light-pulse elastic scattering. Applied to atom interferometry as a study case, this simulator solves the atom-light diffraction problem in the elastic case i.e.\ when the internal state of the atoms remains unchanged. Taking this perspective, the light-pulse beam splitting is interpreted as a space- and time-dependent external potential. In a shift from the usual approach based on a system of momentum-space ordinary differential equations, our position-space treatment is flexible and scales favourably for realistic cases where the light fields have an arbitrary complex spatial behaviour rather than being mere plane waves. Moreover, the numerical package we developed is effortlessly extended to the problem class of trapped and interacting geometries, which has no simple formulation in the usual framework of momentum-space ordinary differential equations. We check the validity of our model by revisiting several case studies relevant to the precision atom interferometry community. We retrieve analytical solutions when they exist and extend the analysis to more complex parameter ranges in a cross-regime fashion. The flexibility of the approach, the insight it gives, its numerical scalability and accuracy make it an exquisite tool to design, understand and quantitatively analyse metrology-oriented matter-wave interferometry experiments. 

\end{abstract}
\begin{document}
\flushbottom
\maketitle
\thispagestyle{empty}

\section*{Introduction and motivation}

The commonly used approach for treating light-pulse beam-splitter and mirror dynamics in matter-wave systems consists in solving a system of ordinary differential equations (ODE) with explicit couplings between the relevant momentum states.  

This formulation starts by identifying the relevant diffraction processes and extracting their corresponding coupling terms in the ODE \cite{berman1997atom, meystre2001atom}.
In the elastic scattering case, each pair of light plane waves can drive a set of two-photon transitions from one momentum class $m$ to the next neighboring orders $m \pm 2$. 
The presence of multiple couplings allows for higher order transitions and the system is simplified by choosing a cutoff omitting small transition strengths.
This ODE approach works well for simple cases leading to analytical solutions in the deep Bragg and Raman-Nath regimes \cite{berman1997atom, meystre2001atom}. Using a perturbative treatement, it was generalised to the intermediate, so-called quasi-Bragg regime\cite{muller2008atom}. A numerical solution in this regime has been extended in the case of a finite momentum width \cite{szigeti2012momentum}. In a different approach, Siem\ss~\textit{et al.}\cite{Siemss2020} developed an analytic theory for Bragg atom interferometry based on the adiabatic theorem for quasi-Bragg pulses. Realistically distorted light beams or mean-field interactions, however, sharply increase the number of plane wave states and their couplings required for an accurate description.
The formulation of the ODE becomes increasingly large and inflexible, with a set of coupling terms for each relevant pair of light plane waves.

Here, we take an alternative approach and solve the system in its partial differential equation (PDE) formulation following the Schr\"odinger equation. This time-dependent perspective\cite{Tannor2018} has several advantages in terms of ease of formulation and implementation, flexibility and numerical efficiency for a broad range of cases. Indeed, this treatment is valid for different types of beam splitters (Bloch, Raman-Nath, deep Bragg and any regime in between) and pulse arrangements. Combining successive light-pulse beam-splitters naturally promoted our solver to a cross-regime or universal atom interferometry simulator that could cope with a wide range of non-ideal effects such as light spatial distortions or atomic interactions, yet being free of commonly-made approximations incompatible with a metrological use.

The position-space representation seems underutilised in the treatment of atom interferometry problems in favor of the momentum-space description although several early attempts of using it were reported for specific cases\cite{simula2007atomic, stickney2008theoretical, liu2009numerical, atus2, blakie2000mean}. In this paper, we show the unique insights this approach can deliver, its great numerical precision and scalability and we illustrate our study with relevant examples from the precision atom interferometry field. 

\section*{Theoretical model}
\subsection*{Light-pulse beam splitting as an external potential}

We start with a semi-classical model of Bragg diffraction, where a two-level atom is interacting with a classical light field\cite{berman1997atom, meystre2001atom}. This light field consists of a pair of two counter-propagating laser beams realised by a retro-reflection mirror setup for example. Assuming that the detuning of the laser light $\Delta$ is much larger than the natural line width of the atom, one may perform the adiabatic elimination of the excited state. This yields an effective Schr\"odinger equation for the lower-energy atomic state $\psi(x, t)$ with an external potential proportional to the intensity of the electric field
\begin{align}
     i \hbar \partial_t \psi(x, t)&= \left( \frac{-\hbar^2}{2M}\frac{\partial^2}{\partial x^2} + 2\hbar\Omega\cos^2(kx) \right)\psi(x, t)
\end{align}
with the two-photon Rabi frequency $\Omega=\Omega_0^2/2\Delta$ and wave vector $k=2\pi/\lambda$ in a simplified 1D geometry along the x-direction. For the present study, we consider a $^{87}Rb$ atom that is addressed at the D2 transition with $\lambda=780$ nm resulting in a recoil frequency and velocity\cite{steck2016rubidium} of $\omega_r=\hbar k^2/2m=2 \pi \cdot 3.8$ kHz and $v_r=\hbar k/m=5.9$ mm/s, respectively.

In the context of realistic precision atom interferometric setups, it is necessary to include Rabi frequencies $\Omega(x,t)$ and wave vectors $k(x,t)$ which are space- and time-dependent. This allows to account for important experimental ingredients such as the Doppler detuning or the beam shapes including wavefront curvatures\cite{louchet2011influence, schkolnik2015effect, zhou2016observing}  and Gouy phases\cite{bade2018observation, wicht2002preliminary, wicht2005phase, clade2006precise}. Moreover, this generalisation allows to effortlessly include the superposition of more than two laser fields interacting with the atoms as in the promising case of double-Bragg diffraction \cite{Kueber2016double,ahlers2016double, giese2013double} and to model complex atom-light interaction processes where spurious light reflections or other experimental imperfections are present\cite{gebbe2019twin}.

\subsection*{Atom interferometer geometries}
The light-pulse representation presented in the previous section is the elementary component necessary to generate arbitrary geometries of  matter-wave interferometers operating in the elastic diffraction limit. Indeed, since the atom-light interaction in this regime conserves the internal state of the atomic system, a scalar Schr\"odinger equation is sufficient to describe the physics of the problem in contrast to the model adopted in reference~[\citenum{atus2}].

For example, a Mach-Zehnder-like interferometer geometry can be generated by a succession of $\frac{\pi}{2}-\pi-\frac{\pi}{2}$ Bragg pulses (beam-splitter, mirror, beam-splitter pulses) of order $n$ separated by a free drift time of $T$ between each pair of pulses. In the case of Gaussian temporal pulses, this leads to a time-dependent Rabi frequency
\begin{align}
    \Omega(t)=\Omega_{bs}e^{\frac{-t^2}{2\tau_{bs}^2}}+\Omega_{m}e^{\frac{-(t-T)^2}{2\tau_{m}^2}}+\Omega_{bs}e^{\frac{-(t-2T)^2}{2\tau_{bs}^2}},
\end{align}
where $\Omega_{bs}$, $\tau_{bs}$ and $\Omega_{m}$, $\tau_{m}$ are the peak Rabi frequencies and their respective durations associated to the beam-splitter and mirror pulses, respectively.
We numerically solve the corresponding time-dependent Sch\"odinger equation using the split-operator method \cite{Feit1982} to propagate the atomic wave packets along the two arms. The populations in the two output ports $\ket{+}=\ket{0\hbar k}$ and $\ket{-}=\ket{2n\hbar k}$ are evaluated after the last recombination pulse waiting for a time of flight $\tau_{ToF}$ long enough such that the atomic wave packets spatially separate. They are obtained by the integration 
\begin{align}
    P^{unnormalised}_{\pm}&=\int_{\pm}\mathrm{d}x\;|\psi(x,\tau_{ToF})|^2,
    \label{eq:pos_measurement}
\end{align}
where the integration domains extend over a space interval with non-vanishing probability density of the states $\ket{\pm}$. These probabilities are further normalised to account for the loss of atoms to other parasitic momentum classes 
\begin{align}
    P_{\pm}&=\frac{P^{unnormalised}_{\pm}}{P_{+}^{unnormalised}+P_{-}^{unnormalised}}.
    \label{eq:normalised_population}
\end{align}
Using Feynman's path integral approach, the resulting phase shift between the two arms can be decomposed as\cite{hogan2008light, storey1994feynman}
\begin{align}
    \Delta \phi &= \Delta \phi_{propagation} + \Delta \phi_{laser} + \Delta \phi_{separation}.
\end{align}
The propagation phase is calculated by evaluating the classical action along the trajectories of the wave packet's centers. The laser phase corresponds to the accumulated phase imprinted by the light pulses at the atom-light interaction position and time. Finally, the separation phase is different from zero if the final wave packets are not overlapping at the time of the final beam splitter, $t=2T$.

To extract the relative phase $\Delta\phi$ between the two conjugate ports and the contrast $C$, one can scan a laser phase $\phi_0\in[0,2\pi]$ at the last beam splitter and evaluate the populations\cite{berman1997atom} varying as
\begin{align}
    P_{\pm}&=\frac{1}{2}\left(1\pm C\cos(\Delta \phi + n\phi_0)\right).
    \label{eq:FringeExpression}
\end{align}
The resulting fringe pattern is then fitted with  $\Delta \phi$ and $C$ as fit parameters. This method, analogous to experimental procedures, allows to determine the relative phase modulo $2\pi$. 

\section*{Results}

\subsection*{Raman-Nath Beam Splitter}
The Raman-Nath regime, characterised by a spatially symmetric beam splitting, is the limit of elastic diffraction for very short interaction times of $\tau \ll \frac{1}{\sqrt{2\Omega \omega_r}}$. The dynamics of the system can, in this case, be analytically captured following references\,[\citenum{berman1997atom, meystre2001atom}]
\begin{align}
    |g_n(t)|^2=J_n^2(\Omega t),
\end{align}
where $g_n(t)$ describes the amplitude of the momentum state $\ket{2n\hbar k}$ and $J_n$ the Bessel functions of the first kind. Such experiments are at the heart of investigations as the one reported in reference\,[\citenum{gupta2002contrast}] where a Raman-Nath beam splitter was used to initialise a three-path contrast interferometer offering the possibility to measure the recoil frequency $\omega_r$. 

\begin{figure}[t] 
    \centering
    \includegraphics[scale=1]{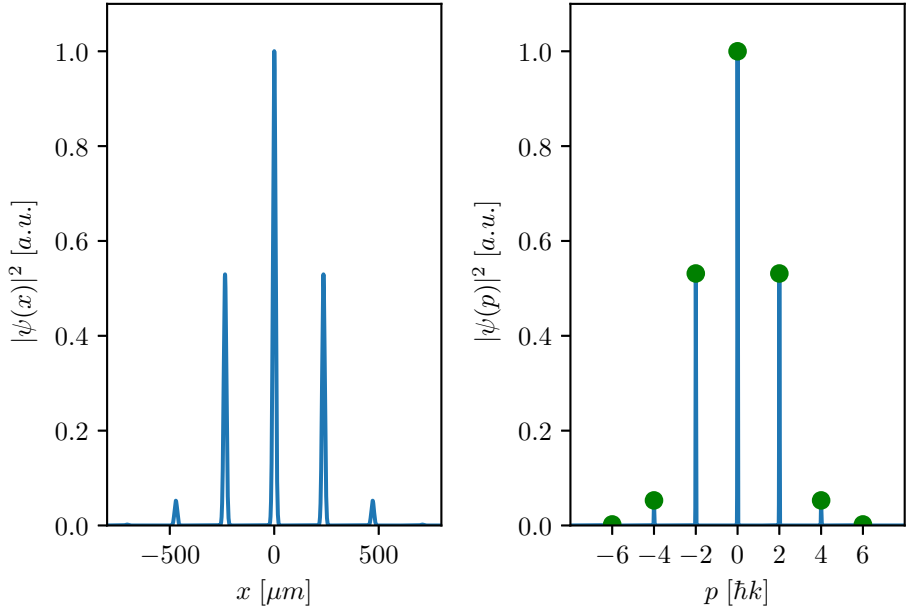}
    \caption{Probability density after a Raman-Nath pulse with $\Omega=50$\,$\omega_r$, $\tau=1$\,$\mu s$ and a rectangular temporal profile as implemented in~[\citenum{gupta2002contrast}]. This shall create a beam splitter of roughly $50\%$ in $\ket{0\hbar k}$ and $25\%$ in each of the $\ket{\pm 2 \hbar k}$ momentum states with an added time of flight of $\tau_{ToF}=20$ ms to clearly separate the wavepackets in position space. The left and right panels show the position- and momentum-space probability density. The initial momentum width of the Gaussian wavepacket is chosen to be $\sigma_p=0.01$ $\hbar$k. Numerical results of this work (continuous blue lines) agree well with the analytical solution of the Raman-Nath regime (green dots, momentum space) given by the Bessel functions of the first kind.}
    \label{fig:raman-nath}
\end{figure}

 To demonstrate the validity of our position-space approach, we contrast our results to the analytical ones obtained adopting the parameters of reference\,[\citenum{gupta2002contrast}]. Fig.~\ref{fig:raman-nath} shows the outcome of a symmetric Raman-Nath beam splitter targeting the preparation of three momentum states: $50\%$ into $\ket{0\hbar k}$ and $25\%$ in each of the $\ket{\pm 2\hbar k}$ momentum classes. As a feature of our solver, we directly observe the losses to higher momentum states ($p=\pm4\hbar k$ and $p=\pm6 \hbar k$) due to the finite pulse fidelity. An excellent agreement is found with the analytical predictions (green filled circles) of the populations of the momentum states. 

\subsection*{Bragg-diffraction Mach-Zehnder interferometers}
To simulate a Mach-Zehnder atom interferometer based on Bragg diffraction, we consider a pair of two counter-propagating laser beams with a relative frequency detuning $\Delta \omega=\omega_1-\omega_2=2 nkv_r$ and a phase jump $\phi_0\in[0,2\pi]$. This gives rise to the following running optical lattice
\begin{align}
     V_{Bragg}(x,t)&= 2 \hbar \Omega(t) \cos^2(k(x-n v_r t) + \frac{\phi_0}{2}).
\end{align}
For sufficiently long atom-light interaction times, i.e.\ in the quasi- and deep-Bragg regimes~\cite{meystre2001atom, keller1999adiabatic, giltner1995theoretical, muller2008atom}, the driven Bragg order $n$ with momentum transfer $\Delta p=2n\hbar k$ is determined by the relative frequency detuning $\Delta \omega$ of the two laser beams. The relative velocity between the initially prepared atom and the optical lattice is $v = nv_r$. In the rest frame of the optical lattice, the atom has a momentum $p=-n\hbar k$. The difference of kinetic energy between the initial ($p=-n\hbar k$) and target state ($p=+n\hbar k$) is vanishing and therefore this transition is energetically allowed and leads to a $\Delta p=n\hbar k- (-n\hbar k)=2n\hbar k$ momentum transfer.

We now realise beam splitters and mirrors by finding the right combination of peak Rabi frequency and interaction time $(\Omega, \tau)$, either by numerical population optimization or analytically, when we work in the deep Bragg regime. Recent advances by Siem\ss~\textit{ et al.}\cite{Siemss2020} generalise this to the quasi-Bragg regime in an analytical description of Bragg pulses based on the adiabatic theorem. For the pulses used in this paper, the two approaches give the same result for the optimized Rabi frequencies and pulse durations. 

\begin{figure} [t]
    \centering
    \includegraphics[scale=1]{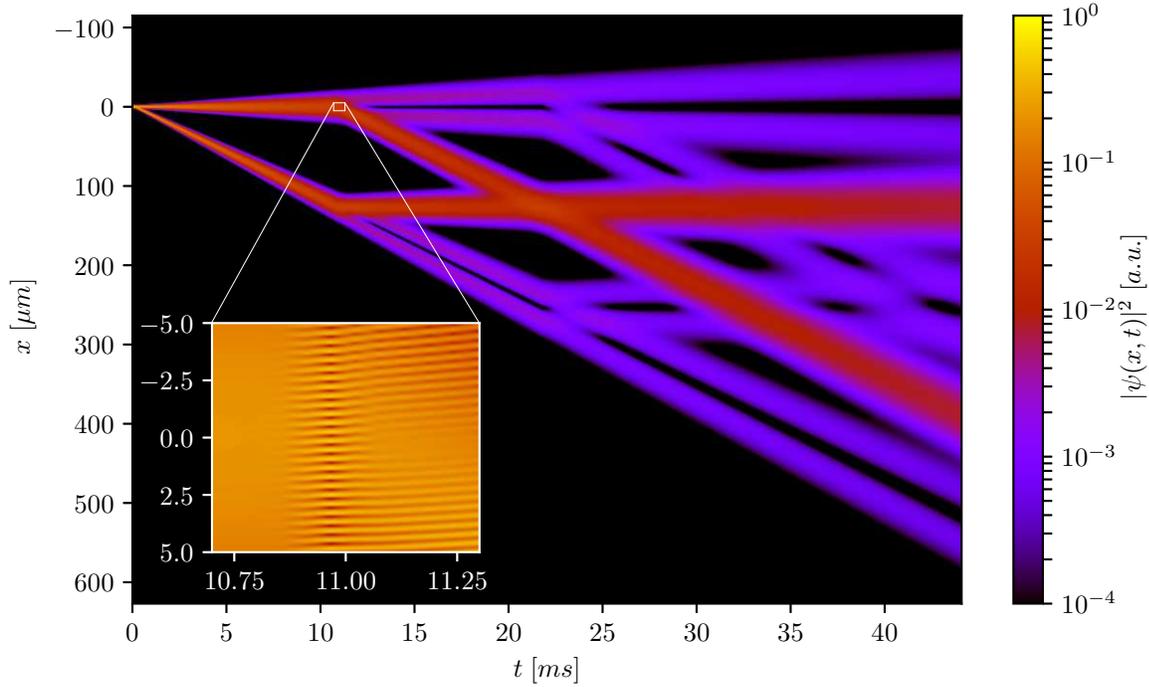}
    \caption{Space-time diagram of the probability density $|\psi(x,t)|^2$ of a $2\hbar k$-Bragg Mach-Zehnder interferometer. The initial momentum width is chosen to be $\sigma_p=0.1$ $\hbar$k and the splitter and mirror Gaussian pulses have peak Rabi frequencies of $\Omega=1.0573$\,$\omega_r$ with pulse lengths of $\tau_{bs}=25$ $\mu$s and $\tau_{m}=50$ $\mu$s, respectively. The separation time between the pulses is $T=10$ ms with a final time of flight after the exit beam splitter of $\tau_{ToF}=20$ ms. Due to the velocity selectivity of the Bragg pulses, several trajectories can be observed after each pulse. \textit{Inset:} Additional insight into the dynamics of the mirror pulse of the upper arm. The peak amplitude of the Gaussian pulse is reached at $t=11$ ms. The interference fringes of the density plot indicate the overlap between the atoms in momentum class $p=2\hbar k$ and the atoms lost due to velocity selectivity remaining at $p=0\hbar k$.}
    \label{fig:MachZehnderBragg}
\end{figure}

In Fig.~\ref{fig:MachZehnderBragg}, we simulate a Mach-Zehnder geometry and illustrate the diffraction outcome by showing a space-time diagram of the density distribution $|\psi(x,t)|^2$. For the parameters chosen here, a clear feature of the dynamics is the appearance of additional atomic channels after the mirror pulse, which can be attributed to the velocity selectivity arising from a pulse with a finite duration characterised by $\tau$. 
The finite velocity acceptance can, indeed, be estimated over the Fourier width $\sigma_f$ of the applied pulse as
\begin{align}
    \mathcal{F}(\Omega e^{-\frac{t^2}{2\tau^2}})=\sqrt{2\pi \tau^2\Omega^2}e^{-2(\pi f\tau)^2},
\end{align}
with $\sigma_f=1/(2\pi \tau)$ and $f$ being the frequency variable. This yields the velocity acceptance\cite{kovachy2012adiabatic} 
\begin{align}
    \sigma_v^{pulse}=\frac{1}{8\,\omega_r \tau}v_r=0.11\,v_r.
\end{align}
With an initial velocity width of the atomic probability distribution of $\sigma_v^{atom}=0.1$\,$v_r,$ it is clear that velocity components with $|v|= \sigma_v^{pulse}$ will have a much smaller excitation probability than the components at the center of the cloud, which leads to the characteristic double well densities of the parasitic trajectories. 

With momenta $p_{upper}=0\hbar k$ and $p_{lower}=2\hbar k$, both parasitic trajectories still fulfill the resonance condition with the final Bragg beam splitter, which leads to the emergence of ten trajectories after the exit beam splitter. For a measurement in position space, it is now important that a sufficiently long time of flight $\tau_{ToF}$ is applied such that the ports of the Mach-Zehnder interferometer do not overlap with the parasitic ports and bias the relative phase measurement.
For large densities, the parasitic trajectories at the Mach-Zehnder ports should not overlap since this may already lead to density interaction phase shifts $H_{int}\propto |\psi(x,t)|^2$. To circumvent these problems it is important to choose $\sigma_v^{pulse} \gg \sigma_v^{atoms}$. 
An example of state-of-the-art experiments\cite{gebbe2019twin} with delta-kick collimated BEC sources\cite{chu_proposal_1986,ammann_delta_1997,morinaga_manipulation_1999,muentinga2013,PRLKovachy2015,NJPCorgier2018} uses $\sigma_v^{atoms}=0.03$\,$v_r\ll0.14$\,$ v_r=\sigma_v^{pulse}$, for strongly suppressed parasitic trajectories due to velocity selectivity. 

\begin{figure} [t]
    \centering
    \includegraphics[scale=1]{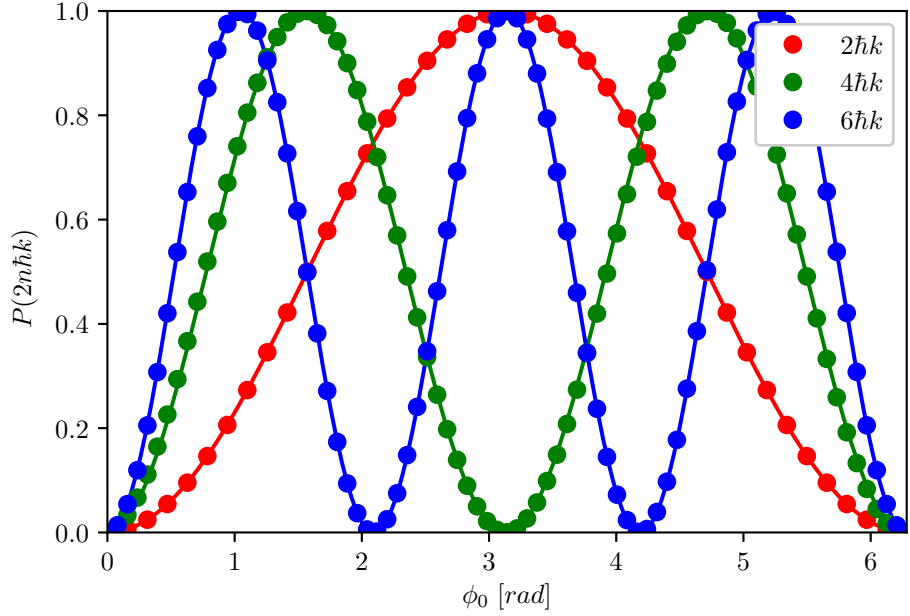}
    \caption{Scan of Mach-Zehnder interferometer phase for different Bragg transition orders of $2\hbar k$ (red dots), $4\hbar k$ (green dots) and $6\hbar k$ (blue dots). The phase shift is applied as a laser phase jump $\phi_0\in[0,2\pi]$ at the last Bragg pulse. The lengths of the Gaussian splitting and mirror pulses are $\tau_{bs}=25$ $\mu$s and $\tau_{m}=50$ $\mu$s, respectively. The initial momentum width of the atomic sample is $\sigma_p=0.01$ $\hbar$k. The corresponding Rabi frequencies for the higher order Bragg transitions were found by optimising for an ideal $50:50$ population splitting of the $\frac{\pi}{2}$ pulse. This leads to $\Omega_{4\hbar k}=3.7$\,$\omega_r$ and $\Omega_{6\hbar k}=8.4$\,$\omega_r$. The Rabi frequency for the $2\hbar k$ transition is $\Omega_{2\hbar k}=1.0573$\,$\omega_r$. The solid lines are the respective fringe scan fits from which the phase shifts and contrasts are directly extracted.}
    \label{fig:higherOrderBragg}
\end{figure}

Implementing high-order Bragg diffraction is a natural avenue to increase the momentum separation of an atom interferometer, and therefore its sensitivity. In Fig.~\ref{fig:higherOrderBragg}, we run our solver to observe the population distribution across of the different ports of a Mach-Zehnder configuration with Bragg orders up to $n=3$. This is done in a straightforward way by scanning the laser phase $\phi_0$. We fit the data points corresponding to the population in the fast port $\ket{2\hbar k}$ for the different Bragg orders according to Eq.~\eqref{eq:FringeExpression} and observe a clear sinusoidal signal of the simulated fringes, as expected. 
The resulting contrasts and phase shifts are directly found by our theory model and numerical solver which include the ideal phase shifts commonly found\cite{storey1994feynman, hogan2008light} and go beyond to comprise several non-ideal effects as \textit{(i)} finite momentum widths, \textit{(ii)} finite pulse timings and \textit{(iii)} multi-port Bragg diffraction\cite{buchner2003diffraction, estey2015high} and the resulting diffraction phase. The natural occurrence of these effects and the possibility to quantify them are a native feature of our simulator.

\subsection*{Symmetric double-Bragg geometry}
Scalable and symmetric atom interferometers based on double-Bragg diffraction were theoretically studied~\cite{giese2013double} and experimentally demonstrated~\cite{ahlers2016double}. This dual-lattice geometry has particular advantages, including an increased sensitivity due to the doubled scale factor compared to single-Bragg diffraction, as well as an intrinsic suppression of noise and certain systematic uncertainties due to the symmetric configuration\cite{ahlers2016double}. Combining this technique with subsequent Bloch oscillations applied to the two interferometer arms, led to reaching momentum separations of thousands of photon recoils as it was recently shown in reference~[\citenum{gebbe2019twin}]. 

\begin{figure}[t]
    \centering
    \includegraphics[scale=1]{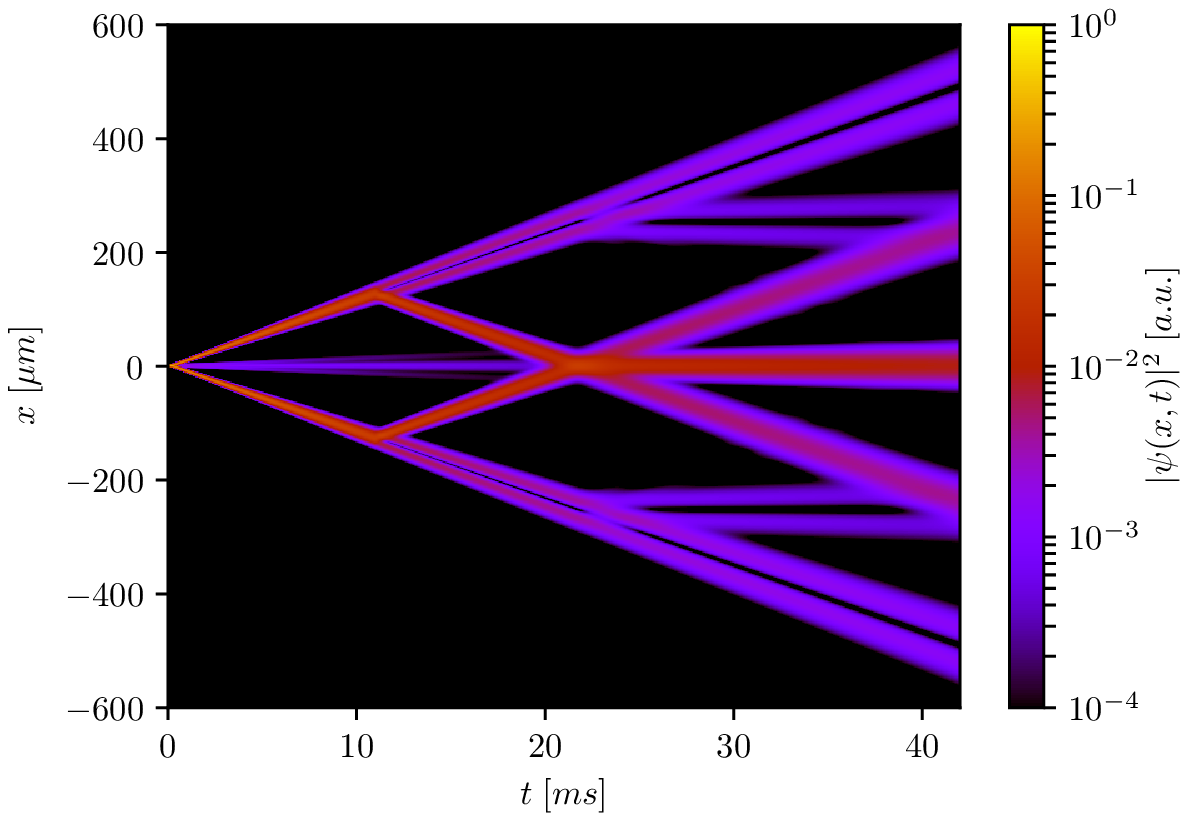}
    \caption{Space-time diagram of a symmetric double-Bragg interferometer. The probability density $|\psi(x,t)|^2$ is plotted for an initial momentum width of $\sigma_p=0.1$ $\hbar$k. The timings of the Gaussian splitter and mirror pulses are set to $\tau_{bs}=25$ $\mu$s and $\tau_{m}=50$ $\mu$s, respectively. The corresponding Rabi frequencies are found by optimising the desired population transfer. The first $\frac{\pi}{2}$ pulse corresponds to a $2\hbar k$ transfer in two directions, realised by two counter-propagating optical lattices which results in a $4\hbar k$ separation between the two interferometer arms. The mirror pulse is a $4\hbar k$ Bragg transition with a Rabi frequency of $\Omega=1.9$\,$\omega_r$ such that both arms make a transition from $\ket{\pm 2\hbar k}\rightarrow \ket{\mp 2\hbar k}$. The last recombination pulse now realises a $50:50$ split of the upper trajectory to $\ket{-2\hbar k}$ and $\ket{0\hbar k}$ and the lower trajectory to $\ket{+2\hbar k}$ and $\ket{0\hbar k}$. This leads to a final population of $25\%$ in the $\ket{\pm2 \hbar k}$ ports and $50\%$ in the $\ket{0\hbar k}$ port. The separation time between the pulses is $T=10$ ms with a final time of flight after the exit beam splitter of $\tau_{ToF}=20$ ms. Due to velocity selectivity of the Bragg pulses and a non-ideal fidelity of the initial beam splitter pulse, several parasitic interferometers can be observed.}
    \label{fig:doubleBragg}
\end{figure}

In double-Bragg diffraction schemes, two counter-propagating optical lattices are implemented such that the recoil is simultaneously transferred in opposite directions, leading to a beam splitter momentum separation of $\Delta p = 4n\hbar k$ \cite{ahlers2016double, giese2013double}. To extend our simulator to this important class of interferometers, we merely have to add a term to the external potential
\begin{align}
     V_{Double\;Bragg}(x,t)&= 2 \hbar \Omega(t) \left(\cos^2(k(x-nv_rt))+\cos^2(k(x+nv_rt)) \right).
\end{align}
The procedures of realising a desired $4n\hbar k $ momentum transfer as well as mirror or splitter pulses are identical to the case of single-Bragg diffraction. A simple scan of the Rabi frequency and pulse timings was enough to obtain a full double-Bragg interferometer as shown in Fig.~\ref{fig:doubleBragg}. The different resulting paths are illustrated in this space-time diagram of the density distribution $|\psi(x,t)|^2$. Similarly to the single-Bragg Mach-Zehnder interferometer we observe additional parasitic interferometers due to the finite velocity filter of the Bragg pulses after the mirror pulse of the interferometer. Due to a finite fidelity of the initial beam splitter, some atoms remain in the $\ket{0\hbar k}$ port and recombine at the last beam splitter with the trajectories of the interferometer. In a metrological study, these effects are highly important to quantify. Our simulator gives access to all the quantitative details of such a realisation in a straightforward fashion.  

\subsection*{Gravity gradient cancellation for a combined Bragg and Bloch geometry}
Precision atom interferometry-based inertial sensors are sensitive to higher order terms of the gravitational potential, including gravity gradients. In particular, for atom interferometric tests of Einstein's equivalence principle (EP), gravity gradients pose a challenge by coupling to the initial conditions, \textit{i.e.} position and velocity of the two test isotopes\cite{Aguilera2014}. A finite initial differential position or velocity of the two species can, if unaccounted for, mimic a violation of the EP. 
By considering a gravitational potential of the form
\begin{align}
    V(x)&=-m g x-\frac{1}{2} m \Gamma x^2,
\end{align}
where $\Gamma=\Gamma_{xx}$ is the gravity gradient in the direction normal to the Earth's surface, the relative phase of a freely falling interferometer can be calculated as\cite{roura2017circumventing}
\begin{align}
    \Delta \phi &= k_{eff}|g-a_{Bragg}|T^2+k_{eff}\Gamma (x_0+v_0T)T^2,
\end{align}
with $k_{eff}=2nk$.

In reference\,[\citenum{roura2017circumventing}], it was shown that introducing a variation of the effective wave vector $\Delta k_{eff}=\Gamma k_{eff}T^2/2$ at the $\pi$ pulse can cancel the additional phase shift due to the gravity gradient. This was experimentally demonstrated in references [\citenum{D'Amico2017,overstreet2018}].

\begin{figure} [t]
    \centering
    \includegraphics[scale=1]{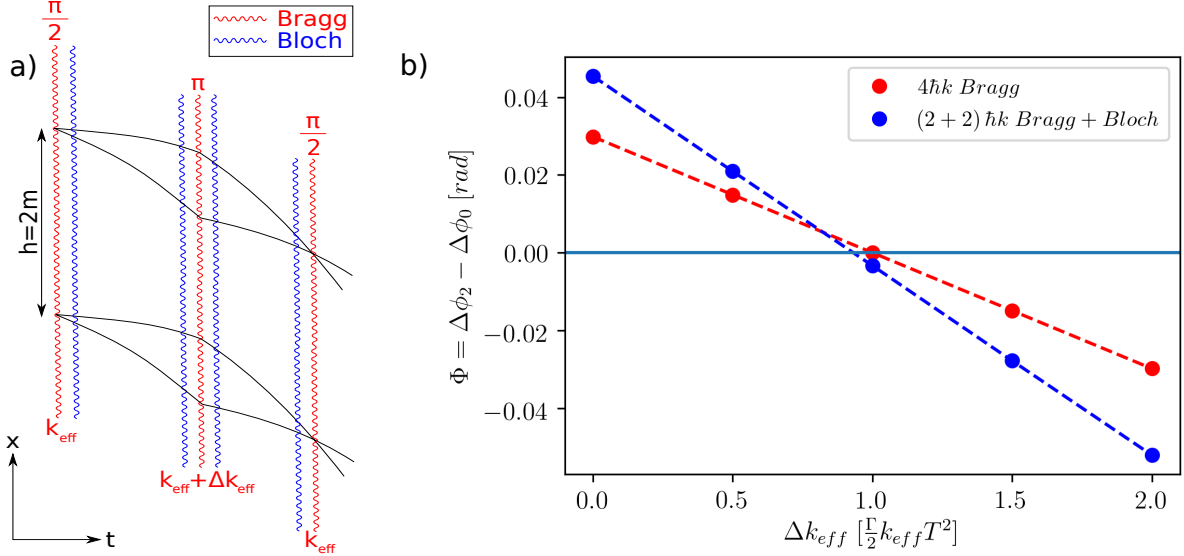}
\caption{Gravity gradient cancellation in the case of a combined Bragg-Bloch gradiometer scheme. a) Schematic of the Bragg-Bloch interferometer geometry with a baseline of $2$\,m. This configuration allows to independently imprint momenta of $n\hbar k$ Bragg and $n_{Bloch}\hbar k$. The Bragg mirror pulse is momentum-adapted to cancel the gravity gradient phase. b) Gradiometric phase for a $4\hbar k$ (red dots) Bragg momentum transfer and a $2\hbar k$  Bragg + $2\hbar k$ Bloch Mach-Zehnder interferometer (blue dots). For both interferometers, the Gaussian pulse lengths of the splitting and mirror pulses are $\tau_{bs}=25$ $\mu$s and $\tau_{m}=50$ $\mu$s, respectively. The initial momentum width of the atomic sample is $\sigma_p=0.01$ $\hbar$k. The corresponding Rabi frequencies for the higher order Bragg transitions were found by optimising for an ideal $50:50$ population splitting of the $\frac{\pi}{2}$ pulse, which leads to a Rabi frequency of $\Omega_{4\hbar k}=3.7$\,$\omega_r$. For the $(2+2)\hbar k$ Bragg+Bloch geometry, the Bloch sequence is implemented with an adiabatic loading time of $\tau_{al}=0.5$ ms, a frequency chirp time of $\tau_{bo}=0.5$ ms during which the momentum transfer occurs and an adiabatic unloading time of $\tau_{aul}=0.5$ ms. The Rabi frequency of the Bloch lattice is $\Omega=4$\,$\omega_r$. For the $4\hbar k$ Bragg geometry we find a vanishing gradiometer phase at $\Delta k_{eff}=\frac{\Gamma}{2}k_{eff}T^2$ which agrees with the analytical calculation\cite{roura2017circumventing}. For the shell $(2+2)\hbar k$  Bragg+Bloch geometry we find a phase shift of $\Phi=-3$ mrad at $\Delta k_{eff}=\frac{\Gamma}{2}k_{eff}T^2$ due to the nontrivial correction of the space-time area  of the  $(2+2)\hbar k$  Bragg+Bloch compared to the $4\hbar k$ Bragg geometry. The dashed lines are a guide to the eye.}
    \label{fig:gravityGradient}
\end{figure}

The same principle applies to the gradiometer configuration of left panel in Fig.~\ref{fig:gravityGradient} where the effect of a gravity gradient is compensated by the application of a wave vector correction. This is reminiscent of another experimental cancellation of the gravity gradient phase shifts\cite{D'Amico2017}. In our example, we first consider a set of two Mach-Zehnder interferometers vertically separated by 2 meters, realised with $4\hbar k$ Bragg transitions where the atoms start with the same initial velocities $v_0$. Choosing a Doppler detuning according to $a_{Bragg}=g$, the gradiometric phase reads
\begin{align}
    \Phi=4k\Gamma h T^2.
\end{align}
By scanning the momentum of the applied $\pi$ pulse, one can compensate the gradiometric phase. This is observed in our simulations at the analytically predicted value of $\Delta k_{eff}=\Gamma k_{eff}T^2/2$ (red dashed curve crossing the zero horizontal line).

It is particularly interesting to use our simulator to find this correction phase in the context of more challenging situations, such as a combined scalable Bragg and Bloch Mach-Zehnder interferometer or a symmetric Bloch beam splitter\cite{Pagel2020} where analytic solutions are not easily found. 

Bloch oscillations can be used to quickly impart a momentum of $p=2n_{Bloch}\hbar k$ on the atoms\cite{dahan1996bloch, wilkinson1996observation}. This adiabatic process can be realised by loading the atoms into a co-moving optical lattice, then accelerating the optical lattice by applying a frequency chirp and finally by unloading the atom from the optical lattice. In our model, this corresponds to the following external potential
\begin{align}
     V_{Bloch}(x,t)&= 2 \hbar \Omega(t)\cos^2(k(x-x(t))) \\
     x(0)&=0\\
     \dot{x}(0)&=2nv_r\\
     \ddot{x}(t)&=
        \begin{cases}
            0, & 0<t<t_{al}\\
            \frac{2n_{Bloch}v_r}{\tau_{bo}} & t_{al}<t<t_{bo}\\
            0 & t_{bo}<t<t_{aul}\\
        \end{cases}\\
     \Omega(t)&=
     \begin{cases}
            \frac{\Omega}{\tau_{al}}t, & 0<t<t_{al}\\
            \Omega, & t_{al}<t<t_{bo}\\
            \Omega\left(1-\frac{t-t_{bo}}{\tau_{aul}}\right) & t_{bo}<t<t_{aul}\\
        \end{cases}
\end{align}
By ramping up the co-moving optical lattice, the atoms are loaded into the first Bloch band with a quasimomentum $q=0$. An acceleration of the optical lattice acts as a constant force on the atoms which linearly increases the quasimomentum over time. When the criterion for an adiabatic acceleration of the optical lattice is met, the atoms stay in the first Bloch band and undergo a Bloch oscillation, which can be repeated $n_{Bloch}$ times leading to a final momentum transfer of $\Delta p=2n_{Bloch}\hbar k$.

The $\pi$ pulse correction $\Delta k_{eff}=\Gamma k_{eff}T^2/2$ is proportional to the space-time area $\mathcal{A}_{Bragg}=\hbar k_{eff}T^2/m$ of the underlying $2n\hbar k$ Mach-Zehnder geometry and does not compensate the gravity gradient effects in the Bloch case. Analysing the space-time area $\mathcal{A}_{Bragg+Bloch}$ immediately shows a non-trivial correction compared to $\mathcal{A}_{Bragg}$. The suitable momentum compensation factor is, however, found using our solver at the crossing of the dashed blue line and the vertical zero limit ($\Delta k_{eff}^{Bragg+Bloch}=0.932$\,$\Delta k_{eff}^{Bragg}$). This straightforward implementation of our toolbox in a rather complex arrangement is promising for an extensive use of this solver to design, interpret or propose advanced experimental schemes.

\subsection*{Trapped interferometry of an interacting BEC}
Employing Bose-Einstein condensate (BEC) sources\cite{ketterle_nobel_2002,cornell_nobel_2002} for atom interferometry\cite{debs2011cold,muentinga2013,Sugarbaker2013} has numerous advantages such as the possibility to start with very narrow momentum widths $\sigma_p$\cite{chu_proposal_1986,ammann_delta_1997,morinaga_manipulation_1999,muentinga2013,PRLKovachy2015,NJPCorgier2018}, which enables high fidelities of the interferometry pulses\cite{szigeti2012momentum}. For interacting atomic ensembles, it is necessary to take into account the scattering properties of the particles. The Schr\"odinger equation is not anymore sufficient to describe the system dynamics and the ODE approach becomes rather complex to use as shown in the section on scalability and numerics.
We rather generalise our position-space approach and consider a trapped BEC atom interferometer including two-body scattering interactions described in a mean-field framework. The corresponding Gross-Pitaevskii equation reads\cite{Pethick2002}
\begin{align}
     i \hbar \partial_t \psi(x, t) &= \left( \frac{-\hbar^2}{2M}\frac{\partial^2}{\partial x^2} + 2 \hbar \Omega(t) \cos^2(k(x-nv_rt)) + g_{1D}N|\psi(x, t)|^2 \right)\psi(x, t),
\end{align}
where the quantum gas of $N$ atoms is trapped is a quasi-1D guide aligned with the interferometry direction and characterised by a transverse trapping at an angular frequency $\omega_{\perp}$. These interactions can effectively be reduced in 1D to a magnitude of $g_{1D}=2\hbar a_{s} \omega_{\perp}$  where $a_{s}=5.272$ nm is the s-wave scattering length of $^{87}$Rb.

All atom interferometric considerations mentioned earlier, like the Bragg resonance conditions, construction of interferometer geometries, the implementation of Doppler detunings, phase calculations and population measurements are also valid in this case without any extra theoretical effort. The non-linear Gross-Pitaevskii equation is solved following the split-operator method as in the Schr\"odinger case\,\cite{Feit1982}.

If the atom interferometer is perfectly symmetric in the two directions of the matterwave guide, no phase shift should occur. In realistic situations, however, the finite fidelity of the beam splitters creates an imbalance $\delta N$ of the particle numbers between the two interferometer arms. The phase shift in this case can be related to the differential chemical potential by
\begin{align}
    \Delta \phi_{MF}&=\frac{1}{\hbar}\int_0^{2T} \mathrm{d}t\;(\mu_{arm1}-\mu_{arm2}).
    \label{eq:meanFieldPhase}
\end{align}
We illustrate the capability of our approach to quantitatively predict this effect by contrasting it to the well-known treatment of this dephasing. Following reference [\citenum{debs2011cold}], we introduce $\delta N \neq 0$ and analyse the dephasing by assuming a uniform BEC density which gives
\begin{align}
    \mu_{arm1/arm2}=\left(\frac{N}{2}\pm\frac{\delta N}{2}\right)\frac{g_{1D}}{2 R_{TF}},
\end{align}
where $R_{TF}$ is the initial Thomas-Fermi radius of the BEC. The $+$ and $-$ signs refer here to the arms 1 and 2, respectively.  Using Eq.~\eqref{eq:meanFieldPhase} we find a phase shift of
\begin{align}
    \Delta \phi_{MF}&=\frac{2T}{\hbar}\left(\frac{\sqrt{m}}{2\sqrt{3}}g_{1D}\omega_x\right)^{2/3} \frac{\delta N}{N^{1/3}}.
\end{align}

\begin{figure} [t]
    \centering
    \includegraphics[scale=1]{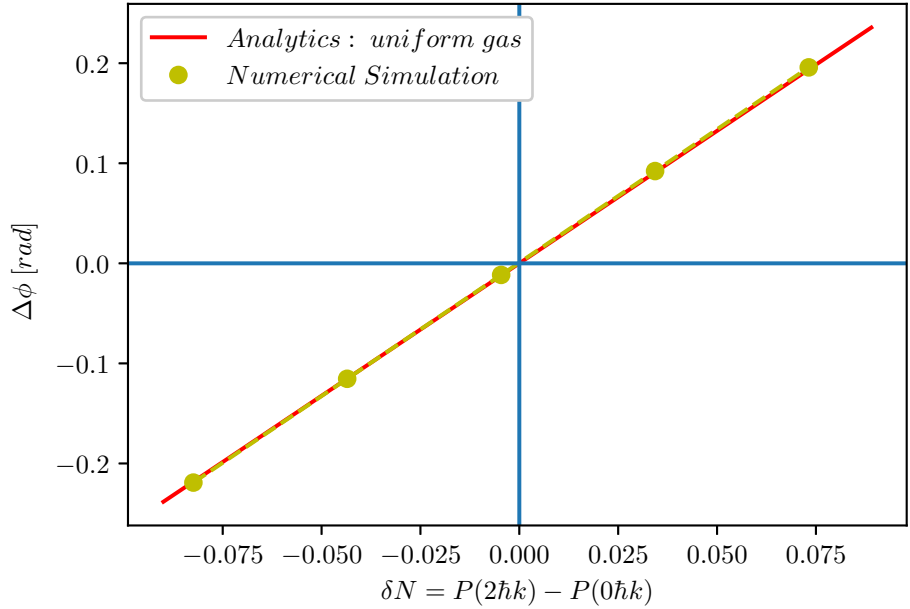}
    \caption{Mean-field-driven phase shifts as a function of the particle imbalance. This imbalance is modeled by considering a first $\frac{\pi}{2}$ beam-splitter with a finite fidelity. The Gaussian splitter and mirror pulses have peak Rabi frequencies of $\Omega=1.0573$\,$\omega_r$ with pulse lengths of $\tau_{bs}=25$ $\mu$s and $\tau_{m}=50$ $\mu$s, respectively. The transverse trapping frequency is realised with an angular frequency of $2\pi \times 50$ Hz and the initial trap frequency in which the BEC is condensed is set to $2\pi\times1$ Hz. The effective 1D scattering length of the atoms is $a_{eff}=10^{-2}$\,$a_s$ with a number of atoms of $N=6\cdot10^4$. At an imbalance of $7\,\%$ we find a phase offset of the numerical results with the uniform density approximation of $2.1$ mrad.}
    \label{fig:MeanFieldResults}
\end{figure}

In Fig.~\ref{fig:MeanFieldResults}, we plot this mean-field shift as a function of the atom number imbalance in the two cases of the numerical solution of the Gross-Pitaevskii equation and with the analytical model using the uniform density approximation. We observe an excellent agreement with a maximal phase shift of $\Delta\phi=2.1$ mrad at an imbalance of $\delta N=7\%$. It is worth noting that the dephasing is accompanied by a loss of contrast consistent with previous theoretical studies\cite{watanabe2012contrast}. We performed a numerical optimisation to find the maximal particle number $N$ up to which we find a contrast of $C>99\,\%$, which is $N\leq6\cdot10^{4}$ in this case. 

\section*{Scalability and numerics}
\label{sec:scalability-numerics}
\subsection*{Numerical accuracy and precision}
\begin{figure} [t]
    \centering
    \includegraphics[scale=1]{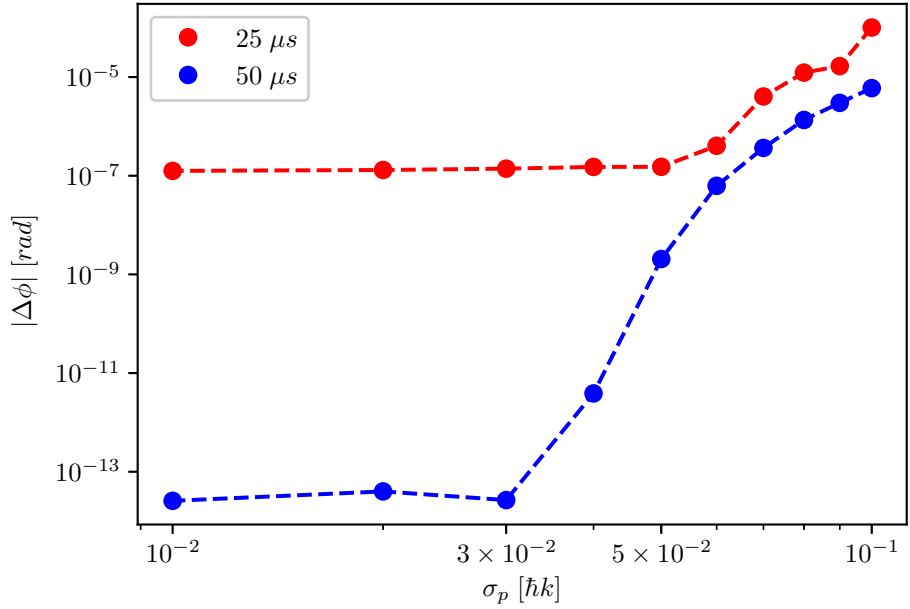}
    \caption{Phase shift offset of a $2\hbar k$ Mach-Zehnder interferometer as a function of the initial momentum width of an atomic sample. We evaluate the phase shift for pulse lengths of $\tau_{bs}=25$ $\mu$s and $\tau_{m}=50$ $\mu$s (blue dots) and for $\tau_{bs}=50 $ $\mu$s and $\tau_{m}=100$ $\mu$s (red dots), using peak Rabi frequencies of $\Omega_{25 \mu s}=1.06$\,$\omega_r$ and $\Omega_{50 \mu s}=0.53$\,$\omega_r$. The dashed lines are a guide to the eye. 
    We find a systematic decreasing behaviour of the relative phase offset $|\Delta \phi|$ starting from an initial momentum width of $\sigma_p=0.1\hbar k$ to $\sigma_p=0.05 \hbar k$ (red dots) and $\sigma_p=0.03\hbar k $ (blue dots). Reaching those critical initial momentum widths both curves show fixed relative phase offsets $|\Delta\phi|$, which in the case of the interferometer with smaller Rabi frequency of $\Omega=0.53\omega_r$ (red dots) reaches a value of $2.5\times10^{-14}$ rad. 
    The numerical simulations were performed with 65536 grid points, an interaction time step of $dt_{int}=1$ $\mu$s and a free evolution time step of $dt_{free}=10$ $\mu$s, leading to a computational time of $\tau_{CPUtime}=12.69$ s on four cores of an Intel Xeon X5670 processor with $2.93$ GHz frequency and $12$ MB of cache.}
    \label{fig:momentumWidthAnalysis}
\end{figure}

To gain a better understanding of the numerical accuracy of the simulations, we plot in Fig.~\ref{fig:momentumWidthAnalysis} the dependency of the phase shift $|\Delta\phi|$ on the momentum width of the atomic sample $\sigma_p$ for a $2\hbar k$ Bragg Mach-Zehnder interferometer. We study two realizations which differ in the peak Rabi frequency with corresponding pulse lengths to perform beam splitter and mirror pulses. For both cases we observe a similar characteristic qualitative behaviour of $|\Delta\phi|$ scaling with $|\Delta\phi|$. Going to smaller initial momentum widths systematically decreases the phase shift until it reaches a plateau of $~1\times 10^{-7}$ rad for $\Omega=1.06\omega_r$ and $2.5\times 10^{-14}$ for $\Omega=0.53\omega_r$. 
 
This qualitative behaviour can be explained by considering the effect of parasitic trajectories. In Fig.~\ref{fig:MachZehnderBragg} it is clearly visible that after the time of flight of $\tau_{ToF}=2T$, there is no clear separation between the parasitic trajectories and the main ports of the Mach-Zehnder interferometer, which leads to interference between them. We choose the integration borders by setting up a symmetric interval around the peak value of each of the ports (cf. Eq.~\eqref{eq:pos_measurement}), ensuring a minimal influence of the parasitic atoms on the interferometric ports. Nevertheless, the interference between the interferometric ports and the parasitic trajectories modifies the measured particle number and therefore also the inferred relative phase. This effect decreases with smaller initial momentum width since less atoms populate the parasitic trajectories overlapping with the main ports, which explains the decrease of relative phase $|\Delta\phi|$ between $\sigma_p=0.1\hbar k$ to $\sigma_p=0.05 \hbar k$ ($\Omega=1.06$\,$\omega_r$) and $\sigma_p=0.03\hbar k $ ($\Omega=0.53$\,$\omega_r$). Another important contribution to the relative phase $|\Delta\phi|$ which is not captured by Feynman's path integral approach\cite{hogan2008light, storey1994feynman} is the diffraction phase, which is fundamentally linked to the excitation of non-resonant momentum states\cite{buchner2003diffraction, estey2015high}. Using smaller Rabi frequencies leads to a reduced population of non-resonant momentum states (after a beam splitter pulse we find $P_{-2\hbar k}^{\Omega=1.06\,\omega_r}+P_{4\hbar k}^{\Omega=1.06\,\omega_r}=1.3\times10^{-7}$ and $P_{-2\hbar k}^{\Omega=0.53\,\omega_r}+P_{4\hbar k}^{\Omega=0.53\,\omega_r}=1.9\times10^{-18}$) and therefore to a reduced diffraction phase which explains that operating a Mach-Zehnder interferometer at $\Omega=0.53$\,$\omega_r$ leads to a much smaller residual phase shift than at $\Omega=1.06$\,$\omega_r$. 

These results indicate that our simulator reaches a relative phase accuracy on the level of $2.5\times10^{-14}$ rad at least. It is worth mentioning, that the numerical parameters chosen to reach this performance are very accessible on modestly powerful desktop computers. 
The computation took $\tau_{CPUtime}=12.7$ s on an Intel Xeon X5670 processor using four cores ($2.93$ GHz, $12$ MB last level cache).  
Modeling precision atom interferometry problems with this method is therefore practical, flexible and highly accurate approach. Using improved resolutions in position and time or higher order operator splitting schemes\cite{javanainen2006symbolic} leads to even better numerical precision and accuracy.

\subsection*{Numerical convergence}
To analyse the numerical convergence of the split-operator method applied to the previously presented systems, we simulate a scalable trapped $(2+2)\hbar k$ Bragg+Bloch transfer for different time and space grids. We consider one Bragg $\pi$ pulse and one Bloch oscillation and quantify the necessary resolution and grid sizes that can be derived from these results.

\begin{figure} [t]
    \centering
    \includegraphics[scale=1]{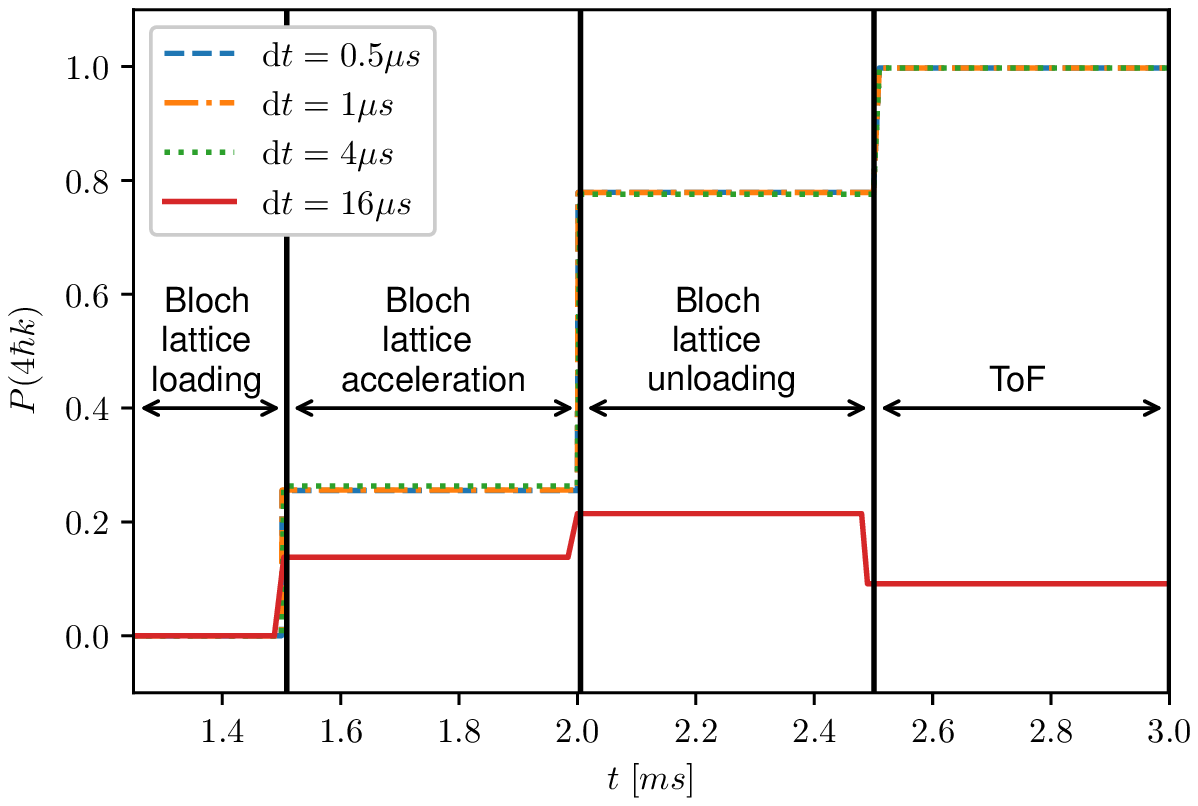}
    \caption{Numerical convergence of a simulated $(2+2)\hbar k$ Bragg+Bloch beam splitter in a waveguide. We analyse the numerical convergence behaviour when changing the time step $\mathrm{d}t$ of the numerical simulation using the third-order split-operator method. The transverse trapping frequency of the waveguide is $2\pi \times 50$ Hz and the initial trap frequency in which the BEC is condensed is set to $2\pi\times1$ Hz. The effective 1D scattering length of the atoms is $a_{eff}=10^{-2}a_s$ with $a_s=5.2$ nm and a number of atoms of $N=6\cdot10^4$. The first Gaussian Bragg pulse has a peak Rabi frequency of $\Omega=1.0573\omega_r$ with pulse length of $\tau=50$ $\mu$s. The following Bloch sequence is implemented with an adiabatic loading time of $\tau_{al}=0.5$ ms, a frequency chirp time of $\tau_{bo}=0.5$ ms during which the momentum transfer occurs and an adiabatic unloading time of $\tau_{aul}=0.5$ ms. The peak Rabi frequency of the Bloch lattice is $\Omega=4\omega_r$. At a spatial step of $\mathrm{d}x=30$ nm, the simulation converges well for a time step of $\mathrm{d}t\approx 1$ $\mu$s.}
    \label{fig:dt_convergence}
\end{figure}

\begin{figure} [t]
    \centering
    \includegraphics[scale=1]{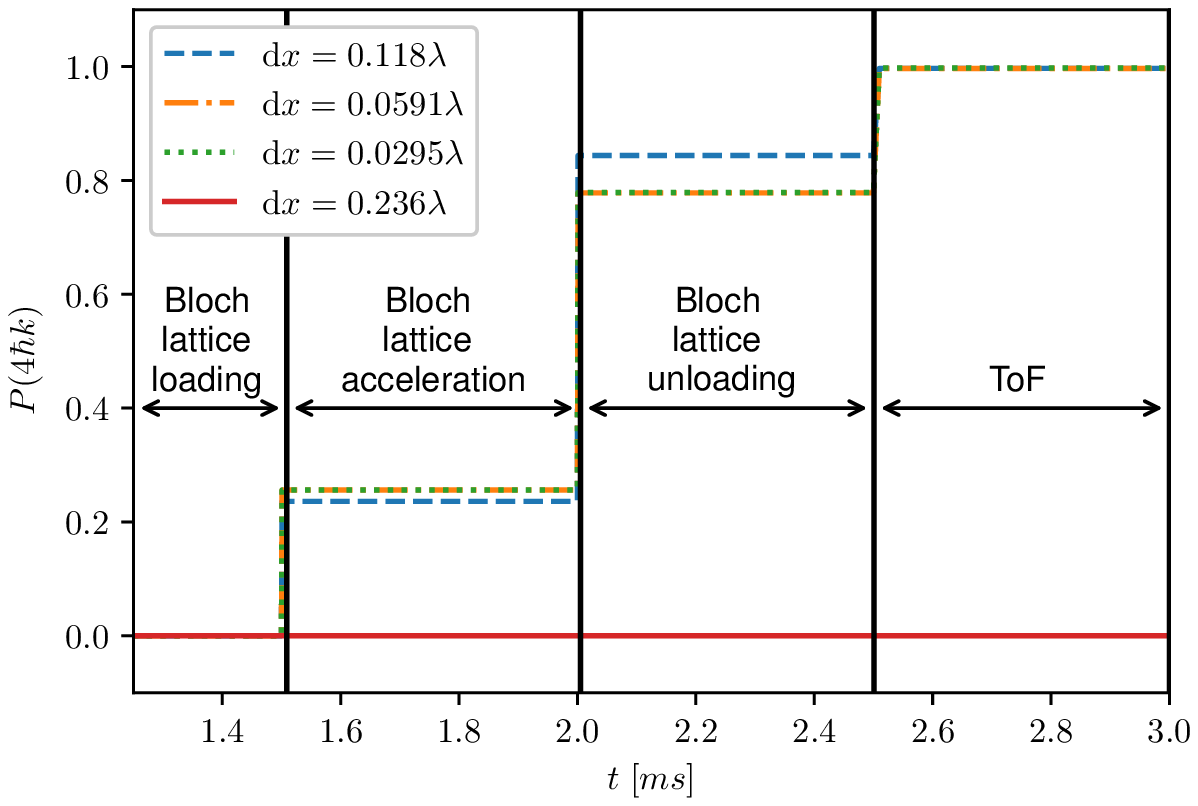}
    \caption{Numerical convergence of a simulated $(2+2)\hbar k$ Bragg+Bloch beam splitter in a waveguide. We analyse the numerical convergence behaviour when changing the position step $\mathrm{d}x$ of the numerical simulation using the third-order split-operator method. The same parameters as Fig.~\ref{fig:dt_convergence} are used. At a time step of $\mathrm{d}t=10^{-6}$ s, the simulation converges well for a position step of $\mathrm{d}x\approx0.06\, \lambda$.}
	\label{fig:dx_convergence}
\end{figure}

Fig.~\ref{fig:dt_convergence} and Fig.~\ref{fig:dx_convergence} show the probability to find an atom in the $4\hbar k$ momentum state $P(4\hbar k)$ after an interaction time $t$. First we drive a Bragg transition to transfer the atoms from the momentum state $p=0\hbar k$ to the state $p=2\hbar k$ and then one Bloch sequence to accelerate the atoms from $p=2\hbar k $ to $p=4\hbar k$. The first step of the Bloch sequence consists of an adiabatic loading of the atoms into the first Bloch state of the lattice for $\tau_{al}=0.5$ ms. This state already has a nonzero $4\hbar k$ momentum component which explains the first jump of the probability $P(4\hbar k)$ at $ t\approx1.5$ ms\cite{peik1997bloch}. The second step of the Bloch sequence is the acceleration of the Bloch lattice to $4$\,$v_r$ where the atoms undergo one Bloch oscillation which is driven at $ \tau_{bo}=0.5$ ms indicated by the second jump of the probability $P(4\hbar k)$. The last step of the Bloch sequence is the adiabatic unloading which converts the atoms of the first Bloch state into the $4\hbar k$ momentum state which can be seen by the last jump at $t\approx2.5$ ms.
 
From Fig.~\ref{fig:dt_convergence} and Fig.~\ref{fig:dx_convergence}, one can extract the critical time and position steps to be $\mathrm{d}t\approx 1$ $\mu$s and $\mathrm{d}x \approx 0.06$\,$\lambda$. These findings can be put into perspective by relating them to the natural time, position, energy and momentum resolutions as well as grid sizes determined from the physical quantities to be resolved. 

The fast Fourier transform (FFT) efficiently switches between momentum and position representations to apply kinetic and potential propagators.
The corresponding position and momentum grids are defined by the number of grid points $N_{grid}$ and the total size of the position grid $\Delta x$ as

\begin{align}
  \mathrm{d}x = \frac{\Delta x}{N_{grid}-1}, \;\;\;\mathrm{d}p = \frac{2 \pi\hbar}{\Delta x}\;\;\; \mathrm{and}\;\;\; \Delta p = \frac{2 \pi \hbar}{\mathrm{d}x},
\end{align}
where $\mathrm{d}p$ and $\mathrm{d}x$ are the steps in momentum and position, respectively, and $\Delta p$ the total size of the momentum grid. 

To resolve a finite momentum width of the atomic cloud we are restricted to 
\begin{align}
  \mathrm{d}p \ll \hbar k \Leftrightarrow \Delta x \gg \lambda,
\end{align}
which sets a bound to the size of the position grid. Finally, $\Delta x$ has to be chosen according to the maximal separation of the atomic clouds $\Delta x_{sep}$. With this we find 
\begin{align}
  \Delta x \gtrsim \Delta x_{sep} \gg \lambda.	 
\label{dp_Deltax}
\end{align}

To include all momentum orders necessary to simulate the considered atom interferometric sequences, we are naturally bound by 
\begin{align}
 \Delta p = \frac{2 \pi\hbar}{\mathrm{d}x} \Leftrightarrow \frac{\Delta p}{\hbar k}=\frac{\lambda}{\mathrm{d}x}
\end{align}
Hence, we find that
\begin{align}
  \frac{\Delta p}{\hbar k}=\frac{\lambda}{\mathrm{d}x} \gg 1 \Leftrightarrow \mathrm{d}x \ll \lambda,
\label{eq:dx_Deltap}
\end{align}
which is the natural condition imposed by the necessity of resolving the atomic dynamics in the optical lattice nodes and anti-nodes of the Bragg and Bloch beams.
 
Fig.~\ref{fig:dx_convergence} shows the limits given by Eq.~\eqref{eq:dx_Deltap}. Choosing $\mathrm{d}x=0.236$\,$\lambda$ leads to a maximal computed momentum of $\pm\,2.1$\,$\hbar k$, which results in the impossibility to find probabilities at $4\hbar k$. Imposing that the position step is roughly one order of magnitude smaller than the wavelength ($\mathrm{d}x \lesssim 0.06$\,$\lambda$) results in a reasonable momentum truncation and in the convergence of the numerical routine.

The typical time scales we need to consider are set on the one hand by the velocities of the optical lattice beams and the atomic cloud and on the other hand by the duration of the atom-light interaction $\tau$. The beams as well as the atomic cloud move with velocities which are proportional to the recoil velocity $v_r$. Given that we want to drive Bragg processes of the order of $n$, we find the following bound on the time step $\mathrm{d}t$
\begin{align}
\mathrm{d}t \ll \frac{\lambda}{nv_r}\approx \frac{100\, \mu s}{n}.
\end{align}
The typical duration of a pulse in the quasi-Bragg regime is strongly depending on the initial atomic momentum width that is being considered. Here, we assume a lower bound of $\tau=10\,\mu s$, which leads to $\mathrm{d}t < \tau$. This bound can be seen in Fig.~\ref{fig:dt_convergence} where the numerical convergence is achieved when we choose a time step $\mathrm{d}t \lesssim 4$ $\mu$s.

It is worth noting that this time step is only necessary during the atom-light interaction. One can simulate the free evolution between the pulses with a much larger time step (without external- and interaction potentials a single step suffices) or using scaling techniques \cite{Castin1996,Kagan1997,vanZoest2010,muentinga2013,meister2017}.

\subsection*{Time complexity analysis}
In this section, we compare the time complexity behaviour of the commonly-used method of treating the beam splitter and mirror dynamics given by the ODE approach to the PDE formulation presented in this paper, based on a position-space approach to the Schr\"odinger equation. To assess the time complexity of the ODE treatment, we re-derive it from the Schr\"odinger equation
\begin{align}
    i \hbar \partial_t \psi(x, t)& = \left( \frac{-\hbar^2}{2M}\frac{\partial^2}{\partial x^2} + 2 \hbar \Omega \cos^2(kx) \right)\psi(x, t).
\end{align}
We decompose the wave function in a momentum state basis as done in references\cite{meystre2001atom, muller2008atom, szigeti2012momentum}
\begin{align}
    \psi(x, t)&=\sum_{m, \delta} g_{m+\delta}(t)\,e^{\frac{i}{\hbar}(m+\delta)px},
\end{align}
where $m$ denotes the momentum orders considered and $\delta$ the discrete representation of momenta in the interval $[k_{m}-k/2,k_{m}+k/2]$ which captures the finite momentum width of the atoms around each momentum class $k_{m}$. Making the two exponential terms appear in $\cos^2(kx)$, one obtains 
\begin{align}
    i \hbar\dot{g}_{m+\delta}(t)&=\hbar\left( (m+\delta)^2 \omega_{r}+ \Omega\right )g_{m+\delta}(t)+\frac{\hbar\Omega}{2}\left(g_{m+\delta+2}(t) + g_{m+\delta-2}(t) \right),
\end{align}
which is a set of $N_{eq}$ coupled ordinary differential equations. This number $N_{eq}$ of equations to solve is equal to $N_{m} N_{\delta}$, set by the truncation condition restricting the solution space to $N_{m}$ momentum classes each discretised in $N_{\delta}$ sub-components. Using standard solvers for such systems like Runge-Kutta, multistep or the Bulirsch-Stoer methods\cite{press1992numerical}, we generally need to evaluate the right hand side of the system of equations over several iterations. With  $N_{eq}$ differential equations, where each one has only two coupling terms, one finds a time complexity of $\mathcal{O}(N_{eq})$.

\begin{figure} [t]
    \centering
    \includegraphics[scale=0.65]{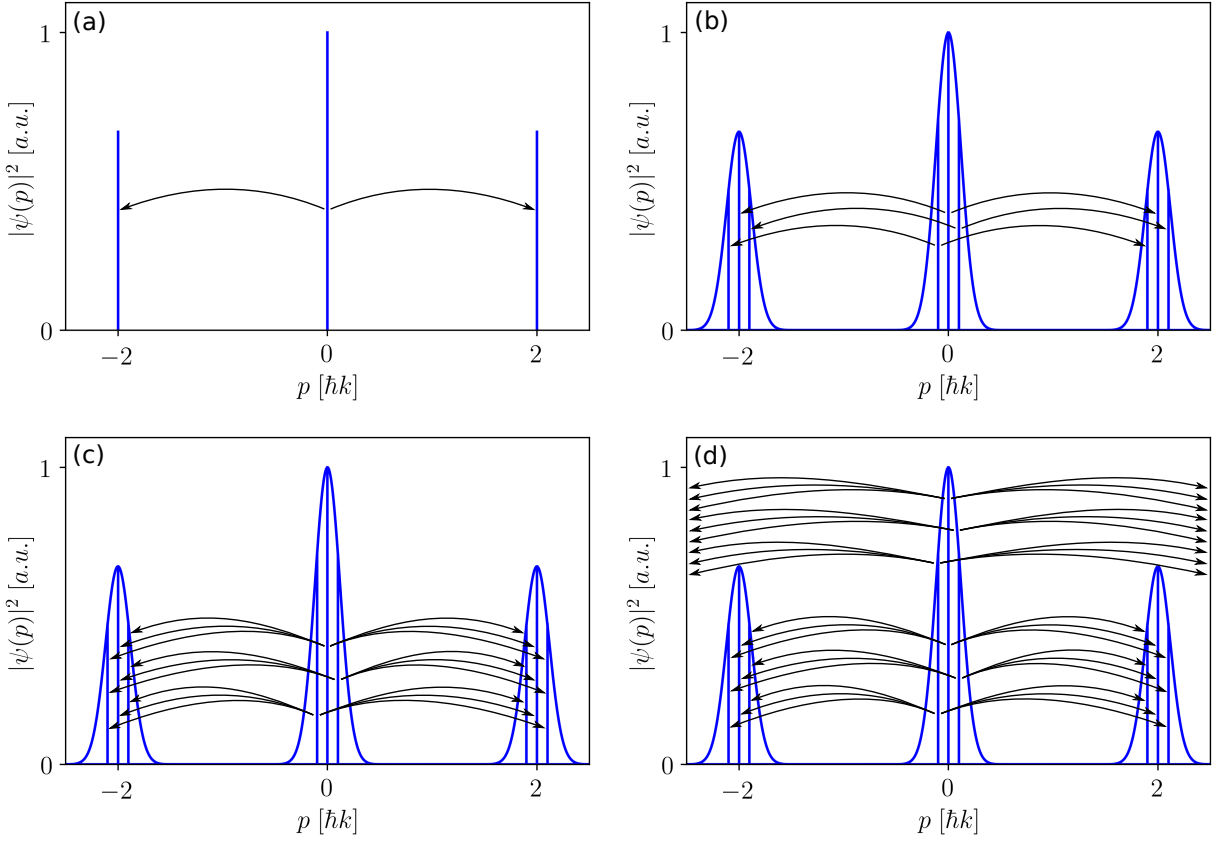}
    \caption{Visualisation of different momentum couplings from the $0\hbar k$ momentum wavepacket corresponding to different levels of complexity. \textit{(a)} Zero momentum width and two coupling elements from $0\hbar k$ to $\pm 2\hbar k$. \textit{(b)} Finite momentum widths with coupling elements for each momentum component in the $0\hbar k$ wavepacket to the corresponding momentum component in the $\pm 2\hbar k$ wavepackets with a momentum difference for each transition of $\Delta p=2\hbar k$. \textit{(c)} Finite momentum widths with multiple possible coupling elements from the $0\hbar k$ wavepacket to the $\pm2\hbar k$ wavepacket, which corresponds to a broadening of the possible momentum difference $\Delta p$. \textit{(d)} Finite momentum widths with higher order coupling elements from the $0\hbar k$ wavepacket to momentum components of the $\pm4\hbar k,\,\pm6\hbar k,\dots$ wavepackets.}
	\label{fig:couplings}
\end{figure}

In a next step, the coupling terms are calculated for more general potentials with time- and space-dependent Rabi frequencies $\Omega(x,t)$ and wave vectors $k(x,t)$. For this purpose, the momentum-space representation of the Schr\"odinger equation is more appropriate and can be written for the Fourier transform of the atomic wave function $g(p,t)$
\begin{align}
    &i \hbar \dot{g}(p,t) = \frac{p^2}{2m}g(p,t)+V(p,t) \ast g(p,t),
\end{align}
where
\begin{align}
    &V(p,t) \ast g(p,t) := \int \mathrm{d}x \;\frac {e^{-i\frac{p}{\hbar}x}}{\sqrt{2 \pi \hbar}}V(x, t)\psi(x, t).
\end{align}
Expressing the wave function in momentum space gives

\begin{align}
    V(p) \ast g(p,t) = \frac{1}{2\pi\hbar}\int \mathrm{d}p'& \underbrace{\int \mathrm{d}x\; e^{ix\frac{(p'-p)}{\hbar}} V(x,t)}_{=:F(p,p',t)\in \mathds{C}} g(p',t).
 \end{align}
Discretising $p \rightarrow (m+\delta)\hbar k$ and $p' \rightarrow (l+\gamma)\hbar k$, one finds   
 \begin{align}
  V(p) \ast g(p,t)  \approx \frac{1}{2\pi\hbar} \sum_{l,\gamma} F( (m+\delta)\hbar k,(l+\gamma)\hbar k,t)g_{l+\gamma}(t),
\end{align}
where $l$ and $\gamma$ span the same indices ensembles as $m$ and $\delta$.
The new equations to solve read 
\begin{align}
i \hbar \dot{g}_{m+\delta}(t) \approx \frac{((m+\delta)\hbar k)^2}{2m}g_{m+\delta}(t) + \frac{1}{2\pi\hbar} \sum_{l, \gamma} F( (m+\delta)\hbar k,(l+\gamma)\hbar k,t)g_{l+\gamma}(t),
    \label{eq:momentumCouplings}
\end{align}
which yields the necessary momentum couplings for an arbitrary potential $V(x,t)$. In the worst case, the sum in Eq.~\eqref{eq:momentumCouplings} runs over $N_{eq}$ nonzero entries ($N_{l} N_{\gamma}=N_{m} N_{\delta}=N_{eq}$) which leads to a time complexity of $\mathcal{O}(N_{eq}^2)$.
This, however, is an extreme example that contrasts with commonly operated precision interferometric experiments since it would correspond to white light with speckle noise. Realistic scenarios rather involve time-dependent potentials with a smaller number of momentum couplings, i.e.\ $N_{eq} \gg \# coupling\; terms \gtrsim 2$. To evaluate the momentum couplings, it is necessary to calculate the integral $F(p,p',t)$ at each time step using the FFT, which leads to a final time complexity class for solving the ODE of $\mathcal{O}(N_{eq}\log N_{eq})$. 

The next important generalisation aims to include the effect of the two-body collisions analysed in the mean-field approximation, i.e.\ $H_{int}=g_{1D}|\psi(x, t)|^2$. In this case, the equation describing the dynamics of the system and the couplings can be written as
\begin{align}
    i \hbar\dot{g}_{m+\delta}(t)=\hbar\left( (m+\delta)^2 \omega_{r} +\Omega\right )g_{m+\delta}(t)&+\frac{\hbar\Omega}{2}\left(g_{m+\delta+2}(t) + g_{m+\delta-2}(t) \right)\\ 
    &+ g_{1D} \left( \sum_{l,\gamma,o,\nu}g^*_{l+\gamma}(t)g_{2o-l+2\nu-\gamma}(t) \right)g_{m+\delta}(t),
\end{align}
where $\nu$ and $o$ are running indices over the same values than $n$ and $\gamma$. One ends up with $N_{eq}$ differential equations where each has more than $N_{eq}^2$ coupling terms, and finds a time complexity class of $\mathcal{O}(N_{eq}^3)$.

In Fig.~\ref{fig:couplings} we see a visualisation of different possible momentum couplings from the $0\hbar k$ component(s) to other momentum components, which capture the different levels of complexity when introducing finite momentum widths, non-ideal light potentials or mean-field interactions. It is clearly visible that the number of coupling elements increases in a non-trivial way starting with only three momentum states and two coupling elements in Fig.~\ref{fig:couplings}~(a) up to more than 9 momentum states and 36 coupling elements in Fig.~\ref{fig:couplings}~(d). This shows the growth in numerical operations of the ODE treatment. Note, that in order to reduce visual complexity we are only showing couplings that start from the $0\hbar k$ wavepacket, while dropping coupling elements starting from $\pm 2\hbar k$. We also fixed the number of momentum states per integer momentum class $k_m$ to three, which in a realistic example is at least an order of magnitude larger. 

We analyse now the time complexity class for the PDE approach, using the split-operator method\cite{Feit1982}. Based on the application of the FFT, it is known that the complexity class of this method is scaling as $\mathcal{O}(N_{grid}\log N_{grid})$,  where $N_{grid}$ is the number of grid points in the position or momentum representations. Since the discretisation of the problem for the ODE and PDE (Schr\"odinger equation) approaches is roughly the same ($N_{eq} \approx N_{grid}$), a direct comparison between the two treatments is possible.

The time complexity analysis is summarised in Table~\ref{table:onlyTableHere}. It shows that the standard ODE approach is only better suited in the case of ideal light plane waves. In every realistic case where the light field is allowed to be spatially inhomogeneous, the amount of couplings increases and it is preferable to employ the PDE approach with a scaling of $\mathcal{O}(N_{grid}\log N_{grid})$, independent of any further complexity to be modelled.
\begin{table}[t]
\centering
\begin{tabular}{ |c||c|c|c| } 
 \hline
 Feature & Numerical Operations ODEs & Numerical Operations PDE \\ 
 \hline\hline
 Infinitely sharp momentum widths ($\delta \rightarrow 1$) & $\mathcal{O}(N_{m})$ & --- \\ 
 \hline
 Finite momentum width and ideal light potential& $\mathcal{O}(N_{grid})$ & $\mathcal{O}(N_{grid}\log N_{grid})$\\
 \hline
 Inhomogeneous light potential & $\mathcal{O}(N_{grid} \log N_{grid}) \rightarrow \mathcal{O}(N_{grid}^2)$ & $\mathcal{O}(N_{grid} \log N_{grid})$ \\
 \hline
 Mean-field interaction & $\mathcal{O}(N_{grid}^3)$ & $\mathcal{O}(N_{grid} \log N_{grid})$ \\ 
 \hline
\end{tabular}
\caption{Comparison of the different time complexity classes of the commonly-used ODE treatment with the position-space approach developed in this work (PDE-based). Including more and more realistic features of the atom-light system leads to an ODE time complexity unfavorably scaling. The PDE formulation, however, routinely scales with $\mathcal{O}(N_{grid}\log N_{grid})$.}
\label{table:onlyTableHere}
\end{table}

\section*{Conclusion}

In this paper, we have shown that the position-space representation of light-pulse beam splitters is quite powerful for tackling realistic beam profiles in interaction with cold atom ensembles. It was successfully applied across several relevant regimes, geometries and applications. We showed its particular fitness in treating metrologically-relevant investigations based on atomic sensors. Its high numerical precision and scalability makes it a flexible tool of choice to design or interpret atom interferometric measurements without having to change the theoretical framework for every beam geometry, dimensionality, pulse length or atomic ensemble property. We anticipate the possibility to generalise this method to Raman or 1-photon transitions if we account for the internal state degree of freedom change during the diffraction. 

\bibliography{main}



\section*{Acknowledgements}

We thank Sven Abend, Sina Loriani, Christian Schubert for insightful discussions and Eric Charron for carefully reading the manuscript. N.G.\@ wishes to thank Alexander D.\@ Cronin for fruitful indications about previous publications related to our current work. 
\\
The presented work is  supported by the VDI with funds provided by the BMBF under Grant No. VDI 13N14838 (TAIOL) and the DFG through CRC 1227 (DQ-mat), project A05 and B07. We furthermore acknowledge financial support from "Niedersächsisches Vorab"
through "Förderung von Wissenschaft und Technik in Forschung und
Lehre" for the initial funding of research in the new DLR-SI Institute and  the “Quantum- and Nano Metrology (QUANOMET)” initiative within the project QT3. Further support was possible by the German Space Agency (DLR) with funds provided by the Federal Ministry of Economic Affairs and Energy (BMWi) due to an enactment of the German Bundestag under grant No. 50WM1861 (CAL) and 50WM2060 (CARIOQA).

\section*{Author contributions statement}

F.F.\@ implemented the numerical model, performed all numerical simulations, and prepared the figures. J.-N.S.\@ and H.A.\@ helped with the interpretation of the results. H.A.\@ and N.G.\@ designed the research goals and directions. E.M.R.\@ and K.H.\@ contributed to scientific discussions. F.F.\@ and N.G.\@ wrote the manuscript. S.S.\@ critically reviewed the manuscript. All authors reviewed the results and the paper and approved the final version of the manuscript.

\section*{Additional information}

Competing interests: The authors declare no competing interests.

\end{document}